\documentclass[aps,prd,twocolumn,superscriptaddress,longbibliography]{revtex4-2}  

\usepackage{graphicx}  
\usepackage[caption=false]{subfig}
\usepackage{dcolumn}   
\usepackage{bm}        

\usepackage{amssymb}   

\usepackage{physics}
\usepackage{xcolor}
\usepackage[normalem]{ulem}

\usepackage{url}\usepackage{hyperref}
\usepackage{bibunits}

\hypersetup{colorlinks,linkcolor={blue!55!black},citecolor={red!50!black},urlcolor={blue!45!black},breaklinks=true}

\def\6{{\langle}}
\def\9{{\rangle}}

\usepackage{dutchcal}

\usepackage{scalerel}
\usepackage{tikz}
\usetikzlibrary{svg.path}
\definecolor{orcidlogocol}{HTML}{A6CE39}
\tikzset{
	orcidlogo/.pic={
		\fill[orcidlogocol] svg{M256,128c0,70.7-57.3,128-128,128C57.3,256,0,198.7,0,128C0,57.3,57.3,0,128,0C198.7,0,256,57.3,256,128z};
		\fill[white] svg{M86.3,186.2H70.9V79.1h15.4v48.4V186.2z}
		svg{M108.9,79.1h41.6c39.6,0,57,28.3,57,53.6c0,27.5-21.5,53.6-56.8,53.6h-41.8V79.1z M124.3,172.4h24.5c34.9,0,42.9-26.5,42.9-39.7c0-21.5-13.7-39.7-43.7-39.7h-23.7V172.4z}
		svg{M88.7,56.8c0,5.5-4.5,10.1-10.1,10.1c-5.6,0-10.1-4.6-10.1-10.1c0-5.6,4.5-10.1,10.1-10.1C84.2,46.7,88.7,51.3,88.7,56.8z};
	}
}

\newcommand\orcidlink[1]{\href{https://orcid.org/#1}{\mbox{\scalerel*{
				\begin{tikzpicture}[yscale=-1,transform shape]
					\pic{orcidlogo};
			\end{tikzpicture}}{X}}}}

\begin{document}

\preprint{APS/123-QED}

\title{Collective Enhancement of Nuclear Excitation for a Nuclear Quantum Battery
}

\author{Pravin Kumar Dahal\,\orcidlink{0000-0003-3082-7853}}
\email{pravin.dahal@csiro.au}
\affiliation{Commonwealth Scientific and Industrial Research Organisation (CSIRO), Clayton, Victoria 3168, Australia}

\author{Kieran Hymas\,\orcidlink{0000-0003-1761-4298}}

\affiliation{Commonwealth Scientific and Industrial Research Organisation (CSIRO), Clayton, Victoria 3168, Australia}

\author{Jack Muir\,\orcidlink{0000-0001-7275-7107}} 

\affiliation{Commonwealth Scientific and Industrial Research Organisation (CSIRO), Clayton, Victoria 3168, Australia}

\author{James Q. Quach} 
\email{james.quach@csiro.au}
\affiliation{Commonwealth Scientific and Industrial Research Organisation (CSIRO), Clayton, Victoria 3168, Australia}

\begin{abstract}

Current implementations of quantum batteries are constrained by limited energy density and short retention times associated with the electronic or molecular excitations. Here we propose a nuclear quantum battery based on collective excitation of the $^{57}$Fe nuclei of density $n$ embedded in a planar hard X-ray waveguide. Using a Green function waveguide-QED description, we study charging via excitation beyond linear response, where saturation and drive back-action reshape the incident pulse. We introduce a self-consistent waveform-engineering protocol that inhibits local radiative decay in the waveguide thus promoting absorption into high-lying collective nuclear excitation manifolds. We show an enhanced excitation cross section of the nuclear ensemble which yields superlinear charging, with maximum studied energy density scaling approximately like $n \sqrt{n}$. Our results provide a route to high-energy-density quantum charging at hard X-ray energies using contemporary X-ray sources and waveguide architectures by identifying nonlinear, collectively enhanced absorption as a key mechanism for nuclear quantum battery operation.

\end{abstract}

\maketitle

\begin{bibunit}[apsrev4-2]

Despite recent progress in quantum battery (QB) platforms, two persistent bottlenecks remain: (i) energy density, set by the energy scale associated with molecular excitations and the achievable excitation fraction, and (ii) retention time, limited by relaxation pathways of the working medium~\cite{Campaioli2023,Quach2020}. QBs are an ensembles of quantum systems designed to store and release energy exploiting the dynamics of the charging process itself, in addition to the material properties of the constituent systems, thereby offering a route to circumvent some of the scaling limitations of conventional devices~\cite{Alicki2013}. In particular, collective charging can enhance the charging power relative to the parallel charging of individual cells, with quantitative bounds and scaling behaviour established in many-body battery models~\cite{Campaioli2017,Ferraro2017,Campaioli2023}. Experimentally, collective charging enhancements consistent with superextensive scaling have been observed in light--matter platforms, highlighting the practical relevance of collectively enhanced absorption processes for QB operation~\cite{Quach2020,Yang2019,Hymas2026}.

To improve both the energy density and retention time of QBs, hybridizing light with nuclear rather than molecular excitations offers a promising approach. Nuclear transition energies are typically in the keV range, far exceeding optical scales, so that even a modest excited-state fraction can correspond to substantial stored energy density. The canonical $^{57}$Fe M\"ossbauer transition at $14.4$~keV exhibits a half-life $\sim 10^2$~ns, corresponding to an extremely narrow nuclear resonance that enables coherent optical control on experimentally accessible timescales~\cite{rohlsberger2021,Bergmann2025,Heeg2021}. These features---together with the prospect of controlling longer-lived nuclear levels, including isomeric transitions in other isotopes, with coherent X-ray fields---motivate nuclear implementations of QB charging protocols beyond the infinitesimal-excitation (linear-response) regime~\cite{Lentrodt2024,rohlsberger2016,wolff2023}.

Nuclear transitions require resonant photons with energies $\sim 10^{4}$ times larger than those of optical transitions. Together with the small single-nucleus radiative cross section, this makes strong excitation of nuclear ensembles challenging without engineered photonic environments~\cite{Lentrodt2024,rohlsberger2016}. Collective emission from nuclear ensembles has been observed in structured X-ray geometries~\cite{Lohse2024,Heeg2021}. However, collectively enhanced absorption, which underpins superextensive charging in QB, has not yet been experimentally demonstrated for nuclear ensembles. X-ray waveguides provide a promising route toward nuclear QBs by enabling controllable light--matter interaction between a nuclear ensemble and a guided electromagnetic mode. Although hard X-ray waveguides are typically weakly guiding and suffer significant cladding absorption, recent experiments have demonstrated resonant propagation and collective nuclear-exciton dynamics, establishing a realistic platform for waveguide QED in the hard X-ray regime~\cite{Lohse2024,Andrejic2023}.

Using a Green function waveguide QED theory for a continuous nuclear ensemble~\cite{Andrejic2023,Lalumiere2013,Sheremet2021}, we establish a collective, superlinear charging of a $^{57}$Fe nuclei embedded in a planar X-ray waveguide which guides magnetic field modes resonant with the $14.4$~keV M\"ossbauer transition. We specifically engineer the incident X-ray waveform to destructively interfere with fields re-radiated locally by the induced nuclear coherence. This suppresses radiative leakage into the guided mode, thereby allowing the ensemble to build up stronger collective correlations and absorb more efficiently in the waveguide. 

\textit{Waveguide nuclear QB--} The $^{57}$Fe M\"ossbauer transition is a magnetic dipole allowed (M1) nuclear transition that transitions the ground $I_g = 1/2$ to the excited $I_e = 3/2$ nuclear spin manifold (here we ignore the dynamics of the $2I + 1$ projective sublevels of each manifold)~\cite{rohlsberger2021}. We treat the nuclei as a continuous ensemble with number density \(\rho(\mathbf r)\) and dimensionless transition operators \(\hat{\sigma}_{ij}(\mathbf r,t)\), such that \(\rho(\mathbf r)\hat{\sigma}_{ij}(\mathbf r,t)\) is the corresponding operator density~\cite{Andrejic2023}. The Hamiltonian is
\begin{equation}
\hat H
=\hbar\omega_0\int d^3r\,\rho(\mathbf r)\,\hat\sigma_{ee}(\mathbf r)
-\int d^3r\,\rho(\mathbf r)\,\hat{\mathbf m}(\mathbf r,t)\cdot \hat{\mathbf B}(\mathbf r,t),
\label{eq:Hcont}
\end{equation}
where \(\omega_0\) is the nuclear transition frequency, \(\hat{\mathbf m}\) is the magnetic dipole operator, and \(\hat{\mathbf B}\) is the magnetic field operator in the structured waveguide environment.

\begin{figure}[!tbp]
    \centering
    \includegraphics[width=0.42\textwidth]{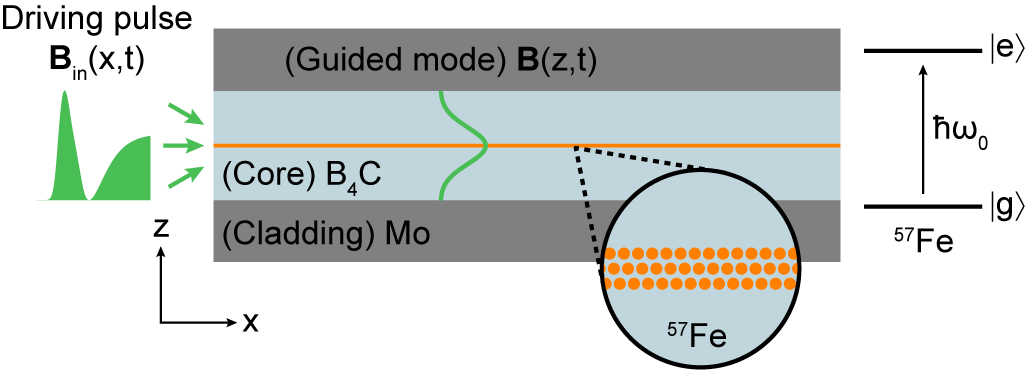}
    \caption{Schematic of a nuclear QB. The device consists of a planar hard X-ray waveguide with a B$_4$C guiding core and Mo cladding. A thin active layer of $^{57}$Fe nuclei is embedded at the centre of the core. The $^{57}$Fe nuclei are modeled as effective two-level systems with ground state $|g\rangle$, excited state $|e\rangle$ and transition frequency $\omega_0$. An incident X-ray field, $\mathbf{B}_{\mathrm{in}}$, driving the nuclear ensemble is shaped to destructively interfere with the field locally re-radiated by the induced nuclear coherence, thereby suppressing radiative leakage into the guided mode and promoting absorption into collective nuclear-exciton manifolds.}
    \label{fig:1}
\end{figure}

In the macroscopic-QED quantization framework, the field in the presence of a polarization source can be written in generalized input--output form as an ``incident'' solution plus the field radiated by the medium-assisted dipole density~\cite{Gruner1996,Dung1997,Asenjo-Garcia2017}.
Specializing to the magnetic field relevant for the M1 M\"ossbauer transition and adopting the continuum ensemble description, we write~\cite{Dung1997,Scheel2008,Andrejic2023}
\begin{multline}
\mathbf{\hat B}_+(\mathbf r,\omega)
= \mathbf{\hat B}_{\mathrm{in}+}(\mathbf r,\omega)
+ \frac{\mu_0\omega^2}{c^2} \\
\int d^3r'\, \rho(\mathbf r')\, \mathbf{G}(\mathbf r,\mathbf r',\omega)\cdot \mathbf{\hat m}_+(\mathbf r',\omega),
\label{eq:fe}
\end{multline}
where $\mathbf{\hat m}_+$ is the positive-frequency component of the magnetic dipole operator. Eliminating the field degrees of freedom yields an effective Hamiltonian describing nuclear interactions mediated by the emission and reabsorption of photons to and from the structured environment. The resulting environment-induced interaction kernel can be decomposed into coherent and dissipative parts~\cite{Lalumiere2013,Sheremet2021}
\begin{equation}
\mathbf{G}(\mathbf r,\mathbf r',\omega)
= \mathbf{J}(\mathbf r,\mathbf r',\omega)
- \frac{i}{2}\,\mathbf{\Gamma}_{\mathrm{rad}}(\mathbf r,\mathbf r',\omega),
\label{eq:Gdecomp}
\end{equation}
where $\mathbf{J}$ generates coherent exchange interactions and $\mathbf{\Gamma}_{\mathrm{rad}}$ is the collective radiative decay-rate kernel (guided plus any other radiative channels).


For a planar waveguide with a single guided mode (see Fig.~\ref{fig:1}), the Green kernel admits a 1D reduction in which propagation along $x$ is mediated by the longitudinal wavevector $k_x$. Adopting the approximation that the guided mode envelope is uniform across the thin resonant layer, we take $\rho(\mathbf r)=L\,\delta(z-z_0)\rho_N(x)$, where $L$ is the layer thickness and $\rho_N(x)$ is the number density within the layer. Under this approximation, substituting Eq.~\eqref{eq:Gdecomp} into Eq.~\eqref{eq:fe} yields
\begin{multline}
\hat B_+(x,z_0,\omega)
= \hat B_{\mathrm{in}+}(x,z_0,\omega)
- i \int dx'\, J(x,x')\, \rho_N(x') \\
\hat{\sigma}_{ge}(x',\omega)
+ \frac{1}{2}\int dx'\, \Gamma_{\mathrm{rad}}(x,x')\, \rho_N(x')\, \hat{\sigma}_{ge}(x',\omega).
\label{eq:B_inout}
\end{multline}
This macroscopic-QED equation for the field inside a waveguide is the starting point for the self-consistent drive engineering developed below. Eq.~\eqref{eq:B_inout} can be written explicitly as (see Supplementary Information Sec. I)
\begin{multline}
\hat B_+(x,z_0,t)
=\hat B_{\mathrm{in}+}(x,z_0,t)
+ \frac{i \mu_0 \omega^2}{c^2} \\
\frac{m_0 L u(z_0)^2}{2 k_x} W_y \int dx'\,\rho_N(x')\,e^{ik_x|x-x'|}\,\hat\sigma_{ge}(x',t),
\label{eq:tfld}
\end{multline}
where $W_y$ is the width of the laser pulse along the y-direction, $m_0$ is the magnetic dipole moment and $u(z)$ is the eigenfunction of the associated Sturm-Liouville problem for TE modes.

The nuclear ensemble interacts with the guided field and evolves according to a master equation in which radiative interactions are mediated by \(\Gamma_{\mathrm{rad}}\) and \(J\), while nonradiative decay enters locally. We write
\begin{equation}
\frac{d}{dt}\hat{\sigma}_{ij}(x,t)
= \frac{i}{\hbar}[ \hat H,\hat{\sigma}_{ij}] + L^H_{\mathrm{nr}}[\hat{\sigma}_{ij}],
\end{equation}
where
\begin{equation}
L^H_{\mathrm{nr}}[\hat O]
= \gamma_{\mathrm{nr}}\int d^3r\,\rho(\mathbf r)\left(
\hat\sigma_{eg}\hat O \hat\sigma_{ge}
-\frac{1}{2}\{\hat O,\hat\sigma_{ee}\}
\right),
\label{eq:Lnr}
\end{equation}
accounts for local, non-radiative dissipation channels such as internal conversion of nuclear excitations at a rate $\gamma_{\mathrm{nr}}$~\cite{Lalumiere2013,Andrejic2023}. 

\begin{figure}[!htbp]
    \centering
    \includegraphics[width=0.42\textwidth]{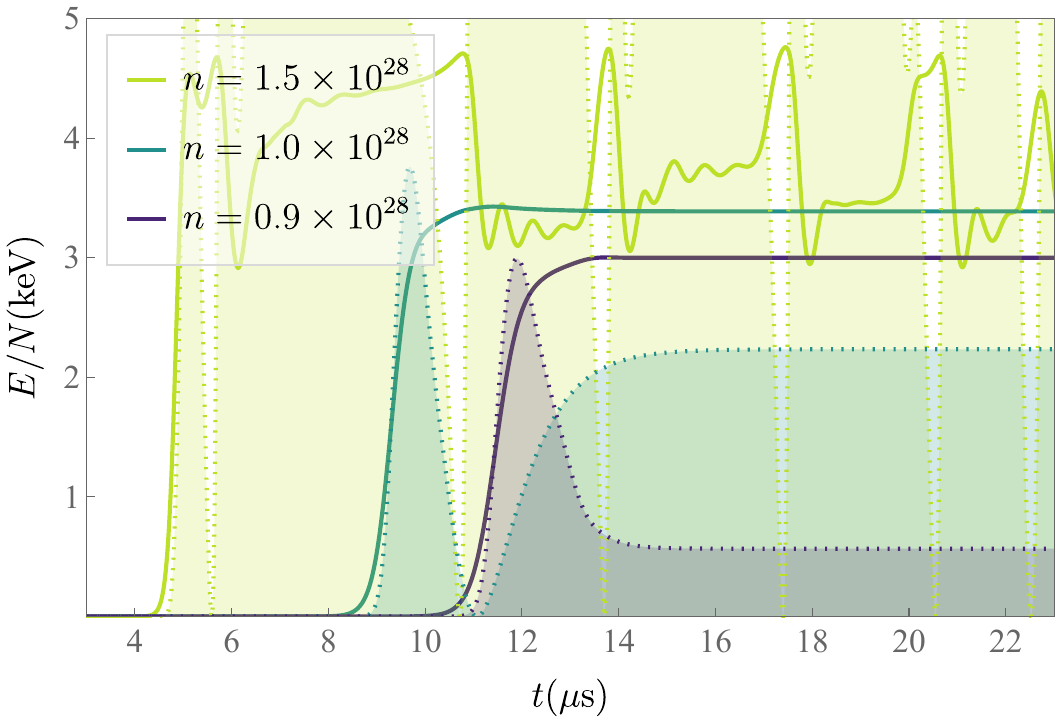}
    \caption{Charging dynamics of nuclear QB. The shaded curves show the optimized incident X-ray waveform used to drive the $^{57}$Fe nuclei in the active layer of the waveguide, for three nuclear number densities: $n=0.9\times10^{28}~\mathrm{m}^{-3}$, $1.0\times10^{28}~\mathrm{m}^{-3}$ and $1.5\times10^{28}~\mathrm{m}^{-3}$. For each $n$, the incident waveform is shaped to destructively interfere with the field locally re-radiated by the induced nuclear coherence, while keeping the total incident energy per nucleus fixed. The solid curves show the corresponding total energy  per nucleus stored in nuclear excitations as a function of time. As $n$ increases, the collective coupling mediated by the guided mode becomes stronger: the excitation rises more rapidly and develops more pronounced oscillations about the quasi-steady plateau.
    }
    \label{fig:3}
\end{figure}

Defining the Bloch operators  $\hat{\sigma}_{ee}= (1+ \hat w)/2$, and $\hat{\sigma}_{ge}= (\hat u+ i \hat v)/2$, and working in the rotating frame at \(\omega_0\), we obtain the Maxwell--Bloch equations (see Supplementary Information Sec. II)
\begin{equation}
    \begin{aligned}
        \partial_t \hat w=& -\left(1+ \hat  w\right) \gamma_{\mathrm{nr}}- \frac{i}{2\hbar} \left(\hat  u (\Omega^*-\Omega)- i \hat  v (\Omega^*+\Omega) \right),\\
        \partial_t \hat u=& -2\hat u \gamma_{\mathrm{nr}}+ \Delta \hat v+ \frac{i}{2\hbar} \hat w (\Omega^*-\Omega),\\
        \partial_t \hat v=& -\hat v \gamma_{\mathrm{nr}}- \Delta \hat u+ \frac{1}{2\hbar} \hat w (\Omega^*+\Omega),
    \end{aligned}
\end{equation}
where $\Delta$ denotes the detuning of the laser frequency and the Rabi frequency is defined as $\Omega= -2 \mathbf{m}_0 \cdot \mathbf{\hat{B}} e^{-i\omega_0 t}\approx -2 \mathbf{m}_0 \cdot \mathbf{\hat B}_-$. In writing these equations, we have used the rotating wave approximation (the validity of approximations used in our model is discussed in Supplementary Information Secs.~I and II).

We consider a hard X-ray waveguide with a thin layer of $^{57}$Fe embedded at the centre of its core (see Fig.~\ref{fig:1}). 
As the X-ray waveguides are weakly guiding (due to the smaller contrast in the refractive index between the core and the cladding) the Purcell enhancement of the X-ray waveguide is negligible~\cite{Lohse2024}. So, we do not rely on discrete-mode cavity strong coupling or vacuum Rabi splitting; instead, the waveguide provides a well-defined one-dimensional propagation channel whose Green’s function mediates long-range interactions between different positions in the nuclear layer, leading to direct nuclear coherent control over interacting modes. For planar waveguides, the dimensionless coupling coefficient over the waveguide modes $\zeta:= W_y L u(z_0)^2\approx 1$ and $k_x= {\mathcal n} \omega/c$, where $\mathcal n$ is the effective index of refraction of the mode. Refractive index of constituent materials in the hard X-ray regime can be written as ${\mathcal n}= 1- \delta+ i\beta$. Here, we take $\delta_{\rm core}= 0.8\times 10^{-6}$, $\delta_{\rm clad}= 3.5\times 10^{-6}$ and $\beta_{\rm core}= 0.5\times 10^{-8}$, $\beta_{\rm clad}= 2.0\times 10^{-8}$ corresponding to the B$_4$C core and Mo cladding, respectively. The core thickness, $d_{\rm core}=18$~nm, is chosen such that the waveguide supports only a single guided mode resonant with the $^{57}$Fe M\"ossbauer transition.

\begin{figure*}[!htbp]
\centering

\subfloat[\label{2A}]{\includegraphics[width=5.77cm]{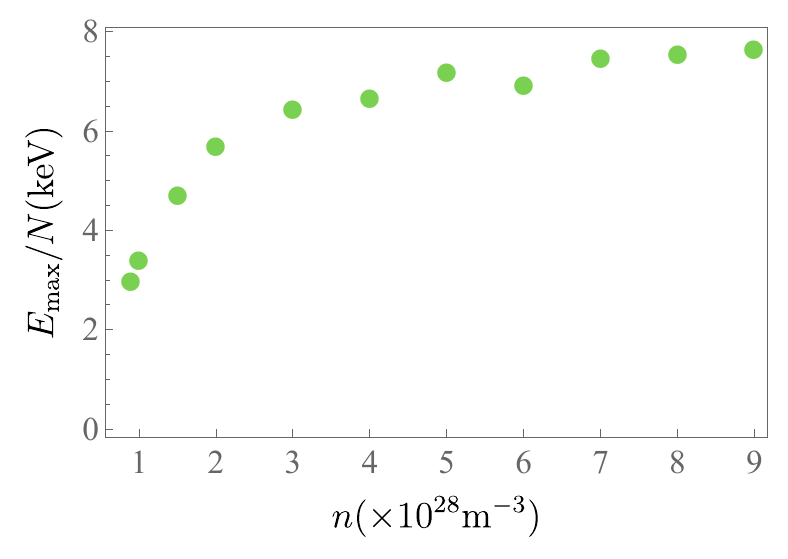}}\quad
\subfloat[\label{2B}]{\includegraphics[width=5.75cm]{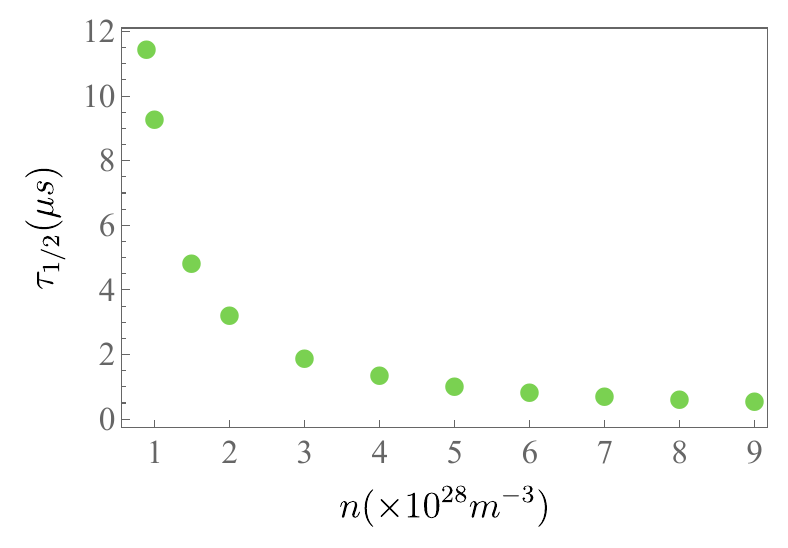}}\quad
\subfloat[\label{2C}]{\includegraphics[width=5.75cm]{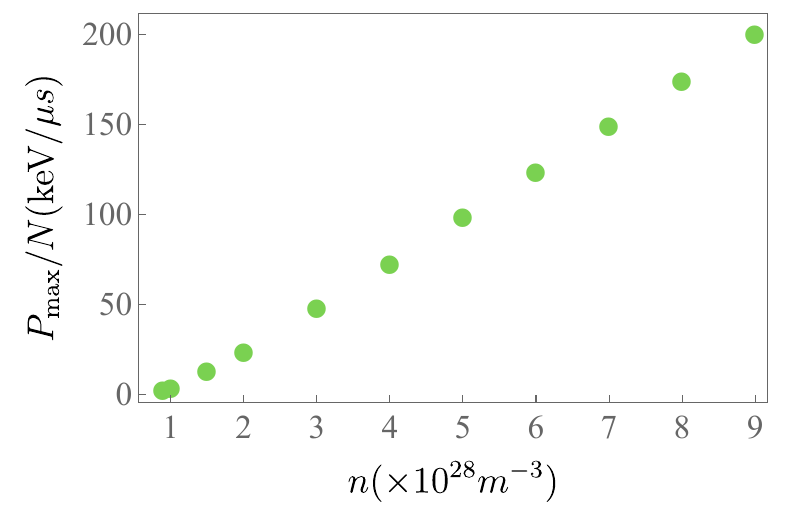}}
        
\caption{Superlinear scaling of nuclear QB charging. (a) Peak stored energy per nucleus as a function of nuclear number density $n$. The increase in peak stored energy with $n$ indicates collectively enhanced absorption in the guided-mode geometry. (b) Charging time, defined as the minimal time required for the stored energy to reach half of its peak stored energy. The charging time decreases with increasing $n$, showing that stronger collective coupling accelerates the charging process. (c) Peak charging power per nucleus, obtained from the largest instantaneous increase in stored energy during the drive. The peak power increases with $n$, consistent with the faster charging time in (b) and the larger peak stored energy in (a). Together, the panels show that increasing $n$ increases energy storage, reduces the charging time and increases the charging power through collective absorption in the guided-mode geometry.}
\label{fig:2}
\end{figure*}

\textit{Superextensive charging--} Direct resonant excitation of nuclei in free space ($\Delta = 0$) using a simple Gaussian pulse requires extremely large peak intensities, which can be reduced by exploiting collective charging advantages in a waveguide geometry. Embedding the nuclear layer within an X-ray waveguide qualitatively alters the charging dynamics by enabling long-range photon-mediated interactions through the guided resonant mode and by enforcing a well-defined propagation geometry for the scattered field. As all nuclei couple to the same guided field, the relevant excitations are collective (Dicke-like) modes rather than independent single-nucleus excitations~\cite{Sheremet2021}. In such collective manifolds, multiple absorption pathways can interfere constructively, leading to enhanced absorption rates compared with an incoherent sum of independent absorbers. 
In suitably engineered many-emitter systems, absorption rates can scale superlinearly with emitter number, ideally approaching $N^2$ under appropriate symmetry and pathway engineering~\cite{Higgins2014}.

A central obstacle to realizing this collective charging advantage is that the same coherence required for enhanced absorption also emits collectively into the guided mode~\cite{Shan2026}, producing a local radiative self-term that represents coherent radiative leakage during the drive. Achieving collective enhancement therefore requires precise preparation of the relevant collective manifold and suppression of premature radiative leakage during the drive, a challenge that becomes acute at hard X-ray energies where mode control is constrained~\cite{Lohse2024}. In thin-film X-ray
cavities, focused-pulse excitation and cavity design have been shown to substantially boost nuclear excitation~\cite{Lentrodt2024}. In an open photonic structure, we instead engineer the incident coherent
drive so that, during the drive, the field re-radiated by the local self-term interferes destructively with the incoming field. This suppression of effective local radiative leakage during preparation increases the available interaction time for building long-range correlations mediated by the guided field, which is essential for observing collective absorption enhancements. Waveform engineering is therefore relevant in the hard X-ray regime, where confinement and mode control are challenging, and strong excitation generally benefits from carefully engineered spatiotemporal driving.

We engineer the charging dynamics by decomposing the radiative decay kernel
\begin{equation}
\Gamma_{\mathrm{rad}}(x,x')
=\gamma_{\mathrm{loc}}(x)\,\delta(x-x')+\widetilde{\Gamma}(x,x'),
\label{eq:Gamma_split}
\end{equation}
into a local radiative decay density $\gamma_{\mathrm{loc}}(x)\,\delta(x-x')$ and a collective, nonlocal radiative decay $\widetilde{\Gamma}(x,x')$ mediated by the guided mode~\cite{Lalumiere2013}. Under this decomposition, the dissipative term in Eq.~\eqref{eq:B_inout} contains an explicit local self-term proportional to the coherence $\tfrac{1}{2}\gamma_{\mathrm{loc}}(x)\,\rho_N(x)\,\hat\sigma_{ge}(x,\omega)$. Our strategy is to shape the incident coherent field such that, during state preparation, it interferes destructively with the locally re-radiated contribution
\begin{equation}
B_{\mathrm{in}+}(x,t)\;\mapsto\;
B_{\mathrm{in}+}(x,t)+\frac{1}{2}\gamma_{\mathrm{loc}}(x)\,\rho_N(x)\,\hat\sigma_{ge}(x,t).
\label{eq:driveEngineering}
\end{equation}
Eq.~\eqref{eq:driveEngineering} constitutes a self-consistent waveform-design, that is, the required incident field depends on the coherence $\hat\sigma_{ge}(x,t)$ that the same drive generates. It represents an impedance-matching condition formulated in real space within the input-output theory for waveguide QED and is closely related to time-reversal mode matching~\cite{Stobinska2009, Leong2016}. More generally, it reflects the broader waveguide QED principle that absorption in a one-dimensional channel depends sensitively on the temporal profile of the incident field~\cite{Yang2019,Chen2011}. Such an engineered coherent drive (self-consistent waveform) enhances absorption into the desired collective manifold while suppressing premature radiative leakage during preparation~\cite{Higgins2014,Sheremet2021}. For each ensemble density $n$, we optimize the incident temporal mode under an
equal charging-resource constraint $\int_0^{t_{\mathrm{End}}}|\Omega_{\mathrm{in}}(t)|^2dt/n$, and determine the corresponding self-consistent waveform over the interval $[0,t_{\mathrm{End}}]$. Importantly, this optimization is defined operationally by the charging-resource constraint and does not rely on choosing a particular reference waveform.

In Fig.~\ref{fig:3}, we quantify charging by the total energy stored in nuclear excitations $E= \hbar\omega_0 n_e(t)$ as a function of time, where $n_e(t)= \int dx\,\rho(x) \hat\sigma_{ee}(x,t)$. We choose $t_{\mathrm{End}}=100/\gamma_{\mathrm{nr}}$ as a representative time-interval to capture the relevant dynamics induced by the self-consistent pulse shaping protocol for different $n$. The optimized
waveform yields a density-dependent charging onset: $E$ rises rapidly and then approaches a quasi-steady plateau while the drive is applied. The oscillations about this plateau, that are particularly evident for $n=1.5\times10^{28}$, arise from collective feedback mediated by the guided field. After the drive ends, the excitation gradually relaxes back at approximately natural decay rate (see Supplementary Information Fig.~S1).

The resulting charging advantage is summarized in Fig.~\ref{fig:2}. Specifically, peak stored energy per nucleus $E_{\max}/N= \hbar\omega_0 n_{e,\textrm{max}}/n$, where $n_{e,\textrm{max}}= \max_t \int dx\,\rho(x) \hat\sigma_{ee}(x,t)$, scales superlinearly with the density $n$ (and thus with total number of nuclei for fixed waveguide dimension). A log--log regression yields $n_{e,\textrm{max}}\propto n^{1.37}$, demonstrating a collective enhancement beyond extensive scaling. Such superlinear scaling is consistent with the general expectation that collective light--matter coupling can enhance charging in QBs, with the attainable exponent set by how efficiently the drive populates the desired collective nuclear-exciton manifold and dissipation in the open waveguide environment. We demonstrate robust superlinear charging in a hard X‑ray nuclear waveguide platform using engineered coherent waveform. Superlinear charging appears only when the ensemble is embedded in a guided mode environment that mediates long-range interactions. In free space, the same driving resources yield smaller excitation and extensive scaling with the nuclei number.

To isolate the role of the waveguide environment, we benchmark the engineered charging protocol against a free-space reference obtained by removing the waveguide Green function and using free-space radiative decay parameters, while retaining nonradiative channels (internal conversion) as local Lindblad terms. In this case, the emitters are effectively independent (no guided mode mediated long-range coupling). For the identical engineered drive envelope $\Omega_{\mathrm{in}}(t)$ applied over the same preparation window $[0,t_{\mathrm{End}}]$, the peak excitation in free space $n_{e,\max}^{\rm fs}$ remains substantially smaller, whereas the waveguide-coupled system achieves markedly larger excitation. The resulting enhancement factor $\eta=n_{e,\textrm{max}}/n_{e,\max}^{\rm fs}$ reaches values up to $\eta\simeq 56$, demonstrating that the observed superlinear charging originates from the waveguide-mediated environment rather than waveform choice.

\textit{Experimental feasibility--} A nuclear QB offers two key advantages over current state-of-the-art QBs. First, the excitation energy of the $^{57}$Fe nucleus is approximately three orders of magnitude larger than that of the molecular excitations used in current QBs. Thus, for the $\sim 30\%$ excitation achieved in Fig.~\ref{fig:3}, the large nuclear transition energy translates into an approximately two orders of magnitude enhancement in stored peak energy relative to molecular-exciton QB platforms. Second, $^{57}$Fe is a M\"ossbauer nucleus with a comparatively long excited-state lifetime of order $\sim 10^2$~ns. Since the relaxation after the driving pulse proceeds approximately at the natural decay rate, the discharge process remains relatively slow (on the microsecond scale), resulting in a storage lifetime about two orders of magnitude longer than that of molecular-exciton QB  platforms.


The proposed protocol is not limited by the availability of peak X-ray intensity, which has been a major obstacle to nonlinear nuclear excitation. The maximum field intensity used in Fig.~\ref{fig:3} is of order $10^{10}~\mathrm{W/cm^2}$, far below the peak intensity of $\sim 10^{22}~\mathrm{W/cm^2}$ achieved with contemporary hard X-ray focusing techniques~\cite{yamada2024}. The main experimental challenge is instead the generation of the extended, structured temporal envelope required by the protocol~\cite{sutter2016}. Such envelopes can be approximated by programmable pulse trains, as demonstrated by solid-state pulse-picking schemes capable of generating patterned X-ray probe sequences on microsecond to millisecond timescales~\cite{schmidt2025}. Moreover, hard X-ray waveguides coupled to M\"ossbauer nuclei have already demonstrated resonant propagation and collective nuclear-exciton dynamics~\cite{Lohse2024}. Thus, the proposed experiment is feasible in the sense that its key components have already been demonstrated separately; it requires integrating them in a single setup that guides a temporally structured X-ray drive, with sufficient resonant spectral weight at the narrow $^{57}$Fe M\"ossbauer transition.

\textit{Long retention nuclear QB--} The underlying principle of our protocol is more general: it is a waveform-matching strategy for enhancing absorption into a target nuclear manifold in an open guided mode environment by suppressing radiative self-loss during preparation. It can therefore be adapted to other nuclei, including long-lived isomers with multiple decay branches and higher-order multipolarities (like E2 and M2), provided a suitable photonic environment exists to mediate collective couplings and shape the incident temporal mode. Appropriate photonic environment concentrates the nuclear emission and absorption into a small number of well-defined electromagnetic modes, so that (a) the ensemble couples to a common field to develop long-range interactions, and (b) the incoming drive can be mode-matched to the dominant radiative channel. Extending collectively enhanced, low-intensity charging schemes to nuclear isomers--with half-lives spanning many orders of magnitude--is thus a natural route toward long-retention nuclear QBs 
; representative examples are summarized in Supplementary Information Table~S1~\cite{walker1999,walker2023,Lestinsky2016}.

The isomer $^{99\mathrm m}$Tc provides a natural long-retention storage level: it lies at $E_{\rm I}\sim 143$~keV above the ground state and has a half-life $T_{1/2}\simeq 6$~h, decaying predominantly by an isomeric transition~\cite{matsuzaki2024}. This motivates an effective $\Lambda$-type model in which the ground state $|g\rangle$ (in $^{99}$Tc) and the metastable state $|m\rangle$ (in $^{99\mathrm m}$Tc) are connected via an auxiliary excited state $|e\rangle$ with allowed electromagnetic links to both states~\cite{liao2011, gao2026}. A temporally engineered drive can then excite population from $|g\rangle$ to $|e\rangle$, from which a fraction relaxes into the isomeric state $|m\rangle$ through the relevant decay channels. In this way, population can be steered into the metastable state using waveform control, reducing radiative loss during preparation through the intermediate level and enabling charging of the long-lived isomer with weaker peak driving than would require in an open radiative environment. Beyond long retention, a key ingredient for a practical nuclear QB is controlled discharge. Controlled discharge requires an active retrieval step, like coherent driving to a radiative transition, or atomic–nuclear coupling processes such as nuclear excitation by electron capture in suitable plasma environments~\cite{Yuanbin2019}. Waveform engineering can naturally be extended from charging to retrieval, thereby directing and accelerating emission. Waveform-engineered collective charging, combined with isomer triggering, provides a pathway toward nuclear QBs that unite (i) charging through collective quantum advantages, (ii) exceptionally long storage set by the isomer lifetime and (iii) energy extraction via collective quantum advantages.

\textit{Conclusion--} We have developed a nuclear QB architecture based on collective excitation of the $^{57}$Fe M\"ossbauer transition embedded in a planar hard X-ray waveguide using a Green function waveguide-QED. A central result is that coherent waveform engineering can substantially improve charging by suppressing effective radiative self-loss during preparation. The incident drive is shaped to destructively interfere with the locally re-radiated field, thereby redirecting the dynamics toward absorption into collective nuclear-exciton manifolds. We thus obtained superlinear scaling of the peak stored excitation with ensemble density, $n_{e,\textrm{max}}\propto n^{1.37}$, together with large enhancements (up to $\eta\simeq 56$) relative to an otherwise identical free-space benchmark. These results identify collectively enhanced absorption in a guided mode environment as a viable mechanism for QB operation at hard X-ray energies. The present framework also provides a pathway to explore alternative isotopes and transitions, including long-lived nuclear isomers, whose excitation energies and lifetimes can be selected to meet specific requirements on energy density and retention time.

\section*{Acknowledgements}

PKD would like to thank Leon Merten Lohse and Yaniv Kurman for useful discussions. The work of PKD is supported by CSIRO Early Research Career postgraduate fellowship.

\putbib[origin]
\end{bibunit}

\clearpage
\begin{bibunit}[apsrev4-2] \makeatletter \renewcommand{\bibnumfmt}[1]{[S#1]} \renewcommand{\citenumfont}[1]{S#1} \makeatother 
\widetext 
\begin{center} \textbf{\large Supplementary Information: Collective Enhancement of Nuclear Excitation for a Nuclear Quantum Battery} \end{center}

\twocolumngrid

\setcounter{equation}{0}
\setcounter{figure}{0}
\setcounter{table}{0}
\setcounter{page}{1}
\makeatletter
\renewcommand{\theequation}{S\arabic{equation}}
\renewcommand{\thefigure}{S\arabic{figure}}
\renewcommand{\thetable}{S\arabic{table}}

\section{Effective atom-atom Hamiltonian} \label{sec:s1}

We consider polycrystalline nuclear ensemble, where the contributions from the Bragg scattering is negligible. This allows us to model the nuclear distribution inside a waveguide as continuum and the Hamiltonian density of the nuclei field can be written as
\begin{equation}
    {\cal H}= \hbar\omega_0 \hat{\sigma}_{ee}(\mathbf{r})- \mathbf{\hat{m}}(\mathbf{r},t)\cdot \mathbf{\hat{B}}(\mathbf{r},t). \label{eq:ham1} 
\end{equation}

We work in the rotating frame of the nuclei, such that, the transition dipole operator and the field can be decomposed into the sum of positive and negative frequency components as
\begin{align}
    \mathbf{\hat m}(\mathbf{r},t)=& \mathbf{\hat m}_+(\mathbf{r},t) e^{-i\omega_0 t}+ \mathbf{\hat m}_-(\mathbf{r},t) e^{i\omega_0 t}, \\
    \mathbf{\hat B}(\mathbf{r},t)=& \mathbf{\hat B}_+(\mathbf{r},t) e^{-i\omega_0 t}+ \mathbf{\hat B}_-(\mathbf{r},t) e^{i\omega_0 t}. \label{eq:mb}
\end{align}
Here, $\mathbf{\hat m}_+(\mathbf{r},t)= \mathbf{m}_0 \hat{\sigma}_{ge}(\mathbf{r},t)$, where $\mathbf{m}_0$ is the magnetic dipole moment. Conventional technique of canonical quantization does not generalize in a simple way to dispersive media because commutation relations for the field operators are not conserved in such media and one must explicitly account for dissipation-induced quantum noise to preserve those relations. We therefore adopt the macroscopic quantization scheme, where the field operator has a generalized input-output form, with the total guided mode being the sum of the probe and scattered fields
\begin{multline}
    \mathbf{\hat B}_+(\mathbf{r},\omega)= \mathbf{\hat B}_{in+}(\mathbf{r},\omega)\\
    + \frac{\mu_0 \omega^2}{c^2} \int d^3 r' \rho(\mathbf{r}') \mathbf{G} (\mathbf{r},\mathbf{r}',\omega)\cdot \mathbf{\hat m}_+(\mathbf{r}',\omega),
\end{multline}
where $\rho(\mathbf{r}')$ corresponds to the number density of the nucleus. The electromagnetic field operators are expressed in terms of the classical dyadic Green tensor of Maxwell's equations together with a continuum of Bosonic noise operators. The Green's function of a waveguide can be separated into two parts
\begin{equation}
    \mathbf{G}= \mathbf{G}^{guid}+ \mathbf{G}^{leaky},
\end{equation}
which correspond to the propagation of guided and leaky modes. In the hard X-ray regime, the leaky mode contribution is small compared to the guided mode contributions. Due to the correspondingly large bandwidth in momentum space, their contributions result in overall Purcell factor and Lamb shift, and thus could be absorbed in the definition of the transition frequency and decay rate~\cite{Andrejic2023}. So, for one-dimensional waveguides, with only one guided mode with the wave vector $k_x$ along the waveguide, the Green's function can be written as~\cite{Sheremet2021}
\begin{equation}
    G(\mathbf{r},\mathbf{r}',\omega)= \frac{i}{2 k_x} u(z) u(z') e^{i k_x |x-x'|}, \label{eq:gf6}
\end{equation}
where $u(z)$ is the eigenfunction of the associated Sturm-Liouville problem for TE modes, which can be found in for example~\cite{Salditt2008}. We also consider a thin resonant nucleus layer, where the guided mode envelopes are uniform across the layer coordinate, such that
\begin{equation}
    \rho(\mathbf{r})= L \delta(z-z_0) \rho_N(x'),
\end{equation}
where $L$ is the thickness of layer, $z_0$ is the centre of $z$ coordinate and $\rho_N(x')$ is the nuclear number density within the layer. Under this approximation, the guided-mode contribution to the magnetic field is nonlocal only along $x$, and the field equation reduces to
\begin{multline}
    \hat B_+(x,z_0,t)= \hat B_{in+}(x,z_0,t)\\
    + \frac{i \mu_0 \omega^2}{c^2} \frac{m_0 L u(z_0)^2}{2 k_x} W_y\int d x' \rho_N(x') e^{i k_x |x-x'|} \hat{\sigma}_{ge}(x',t),
\end{multline}
where $W_y$ is the width of the laser beam along the $y$-direction. We can furthermore simplify the coefficients to write the field operator as
\begin{multline}
    \hat B_+(x,z_0,t)= \hat B_{in+}(x,z_0,t)\\
    + \frac{i \hbar\Gamma}{4 m_0} \zeta \int d x' \frac{e^{i k_x |x-x'|}}{\Lambda_{res}} \hat{\sigma}_{ge}(x',t), \label{eq:tfldS}
\end{multline}
where $\Lambda_{res}= 1/\rho_N \sigma_{rad}$ is the on-resonance attenuation length and $\sigma_{rad}= 2 m_0^2 \mu_0 k_0/\hbar\Gamma$ is the effective resonant cross-section of the nuclei.

\paragraph{Slowly varying envelope.}

For one-dimensional waveguide, the Green's function of Eq.~\eqref{eq:gf6} can be written schematically
\begin{equation}
    G(x,x',\omega)\propto A(\omega)\,e^{i k_x(\omega)|x-x'|}~,
\end{equation}
where $A(\omega)$ is some slowly varying envelope. Because the $^{57}$Fe M\"ossbauer transition at resonant frequency has extremely narrow linewidths, the relevant frequencies satisfy $\omega=\omega_0+\Delta\omega$ with $|\Delta\omega|\lesssim 1/T_1$. Here $T_1= 1/\Gamma$ is the relaxation time. Over this linewidth, the effective index of refraction associated with the guided mode vary negligibly, so we linearize the wave vector
\begin{equation}
k_x(\omega)\simeq k_x(\omega_0)+(\omega-\omega_0)\left.\frac{\partial k_x}{\partial\omega}\right|_{\omega_0}
= k_{x0}+\frac{\omega-\omega_0}{v_g}, \label{eq:dr11}
\end{equation}
where $v_g^{-1}=\left.\partial k_x/\partial\omega\right|_{\omega_0}$. For weakly guiding hard X-ray waveguide, $v_g\approx c$. Thus, the rapidly oscillating propagating phase can be eliminated by redefining the operators as $\hat O(x,\omega)\to e^{-i\omega x/v_g}\hat O(x,\omega)$, which has the effect of replacing time in the Fourier inversion by a retarded time $t_r=t-x/v_g$. For $T_1\sim 10^2$ ns and propagation distances of $|x-x'|\lesssim 1$ mm, the retardation time is $|x-x'|/v_g \lesssim 3.3~{\rm ps}\approx 3\times 10^{-5} T_1$, hence negligible on the nuclear timescales. Therefore, over the spatial extent relevant here, retardation across the nuclear response can be neglected, and we may set $t-|x-x'|/v_g\simeq t$) throughout our calculations.

We use the dispersion relation of Eq.~\eqref{eq:dr11} to separate the propagator as
\begin{equation}
e^{i k_x|x-x'|}= e^{i k_x(\omega_0)|x-x'|} e^{i(\omega-\omega_0)|x-x'|/v_g}~.
\end{equation}
The variation of the second factor, which is responsible for retardation in the time domain, across the nuclear bandwidth is bounded by
\begin{equation}
\left|\frac{(\omega-\omega_0)|x-x'|}{v_g}\right| \lesssim \frac{|x-x'|}{v_g T_1}.
\end{equation}
Thus, over the spatial extent relevant here, we may set $e^{i(\omega-\omega_0)|x-x'|/v_g}\approx 1$.


\section{Evolution of transition operators}

In components form, the master equation for the $^{57}$Fe nuclei, written in Eq.~(6) takes the form
\begin{equation}
    \begin{aligned}
        \frac{\partial}{\partial t} \hat{\sigma}_{gg}(x,t)=& \gamma_{\mathrm{nr}} \hat{\sigma}_{ee}- \frac{i}{\hbar} \left(\hat{\sigma}_{eg} e^{i\omega_0 t} -\hat{\sigma}_{ge} e^{-i\omega_0 t} \right) \mathbf{m}_0 \cdot \mathbf{\hat{B}}, \\
        \frac{\partial}{\partial t} \hat{\sigma}_{ee}(x,t)=& -\gamma_{\mathrm{nr}} \hat{\sigma}_{ee}+ \frac{i}{\hbar} \left(\hat{\sigma}_{eg} e^{i\omega_0 t} -\hat{\sigma}_{ge} e^{-i\omega_0 t} \right) \mathbf{m}_0 \cdot \mathbf{\hat{B}}, \\
        \frac{\partial}{\partial t} \hat{\sigma}_{ge}(x,t)=& \left(-i \Delta- \frac{\gamma_{\mathrm{nr}}}{2} \right) \hat{\sigma}_{ge}- \frac{i}{\hbar} \left(\hat{\sigma}_{ee} - \hat{\sigma}_{gg} \right) e^{i\omega_0 t} \mathbf{m}_0\cdot \mathbf{\hat{B}}, \\
        \frac{\partial}{\partial t} \hat{\sigma}_{eg}(x,t)=& \left(i \Delta- \frac{\gamma_{\mathrm{nr}}}{2} \right) \hat{\sigma}_{eg}+ \frac{i}{\hbar} \left(\hat{\sigma}_{ee} - \hat{\sigma}_{gg} \right) e^{-i\omega_0 t} \mathbf{m}_0\cdot \mathbf{\hat{B}},
    \end{aligned}
\end{equation}
where $\Delta$ denote the detuning of the laser. Substituting the expression for $\mathbf{\hat{B}}$ from Eq.~\eqref{eq:mb} and using the rotating wave approximation to neglect terms containing $e^{\pm 2 i\omega_0 t}$ (as nuclear transition frequency is large, this approximation holds well), we obtain
\begin{equation}
    \begin{aligned}
        \frac{\partial}{\partial t} \hat{\sigma}_{gg}(x,t)=& \gamma_{\mathrm{nr}} \hat{\sigma}_{ee}- \frac{i}{\hbar} \left(\hat{\sigma}_{eg} \mathbf{m}_0 \cdot \mathbf{\hat B}_+ -\hat{\sigma}_{ge} \mathbf{m}_0 \cdot \mathbf{\hat B}_- \right) , \\
        \frac{\partial}{\partial t} \hat{\sigma}_{ee}(x,t)=& -\gamma_{\mathrm{nr}} \hat{\sigma}_{ee}+ \frac{i}{\hbar} \left(\hat{\sigma}_{eg} \mathbf{m}_0 \cdot \mathbf{\hat B}_+ -\hat{\sigma}_{ge} \mathbf{m}_0 \cdot \mathbf{\hat B}_- \right) , \\
        \frac{\partial}{\partial t} \hat{\sigma}_{ge}(x,t)=& \left(-i \Delta-\gamma_{\mathrm{nr}}/2\right) \hat{\sigma}_{ge}- \frac{i}{\hbar} \left(\hat{\sigma}_{ee} - \hat{\sigma}_{gg} \right) \mathbf{m}_0 \cdot \mathbf{\hat B}_+, \\
        \frac{\partial}{\partial t} \hat{\sigma}_{eg}(x,t)=& \left(i \Delta-\gamma_{\mathrm{nr}}/2\right) \hat{\sigma}_{eg}+ \frac{i}{\hbar} \left(\hat{\sigma}_{ee} - \hat{\sigma}_{gg} \right) \mathbf{m}_0 \cdot \mathbf{\hat B}_-, \label{eq:merwa}
    \end{aligned}
\end{equation}
To simplify these equations for the two-level system, we make the following substitutions
\begin{equation}
    \hat{\sigma}_{ee}= \frac{1+ \hat w}{2}, \quad \hat{\sigma}_{ge}= \frac{\hat u+ i \hat v}{2} .
\end{equation}
We obtain~\cite{mccall1969}
\begin{equation}
    \begin{aligned}
        \partial_t \hat w=& -\left(1+ \hat  w\right) \gamma_{\mathrm{nr}}- \frac{i}{2\hbar} \left(\hat  u (\Omega^*-\Omega)- i \hat  v (\Omega^*+\Omega) \right),\\
        \partial_t \hat u=& -2\hat u \gamma_{\mathrm{nr}}+ \Delta \hat v+ \frac{i}{2\hbar} \hat w (\Omega^*-\Omega),\\
        \partial_t \hat v=& -2\hat v \gamma_{\mathrm{nr}}- \Delta \hat u+ \frac{1}{2\hbar} \hat w (\Omega^*+\Omega),
    \end{aligned}
\end{equation}
where the Rabi frequency is defined as $\Omega= -2 \mathbf{m}_0 \cdot \mathbf{\hat{B}} e^{-i\omega_0 t}\approx -2 \mathbf{m}_0 \cdot \mathbf{\hat B}_-$. 

Since the parameters involved in the calculations are in the extreme scale, we define dimensionless parameters by rationalizing them. We define dimensionless time $\tau= 2 t \gamma_{\mathrm{nr}}$, dimensionless Rabi frequency $\hat\Omega= \Omega/\Omega_0$, dimensionless detuning $\hat\Delta= \Delta/2\gamma_{\mathrm{nr}}$. We also introduce a key dimensionless parameter $S= \Omega_0/2\gamma_{\mathrm{nr}}$, which defines the driving strength. The Bloch variables $\hat u$, $\hat v$ and $\hat w$ are already dimensionless, so they remain unchanged. Dimensionless Bloch equations then becomes
\begin{equation} \label{eq:nucl}
    \begin{aligned}
        \partial_\tau \hat w=& -\frac{1+ \hat  w}{2}- \frac{i S}{2\hbar} \left(\hat  u (\hat\Omega^*- \hat\Omega)- i \hat  v (\hat\Omega^*+\hat\Omega) \right),\\
        \partial_\tau \hat u=& -\hat u+ \hat\Delta \hat v+ \frac{i S}{2\hbar} \hat w (\hat\Omega^*-\hat\Omega),\\
        \partial_\tau \hat v=& -\hat v- \hat\Delta \hat u+ \frac{S}{2\hbar} \hat w (\Omega^*+\Omega).
    \end{aligned}
\end{equation}
We use these set equation for the evolution of the nuclear ensemble and Eq.~\eqref{eq:tfldS} for the evolution of electromagnetic field inside a planar waveguide. Our nonlinear waveguide-QED Maxwell--Bloch model is derived under: 
(i) a rotating-wave approximation at the nuclear carrier frequency; 
(ii) a Markov approximation for the radiative reservoir determining \(\Gamma\) and \(J\); 
and (iii) a one-dimensional guided-mode projection (with any additional radiative channels incorporated in \(\Gamma\) and nonradiative channels treated separately as local Lindblad terms, e.g.\ internal conversion). These assumptions are standard in waveguide-QED input--output treatments and in macroscopic-QED descriptions of resonant M\"ossbauer ensembles in structured X-ray environments~\cite{Lalumiere2013,Andrejic2023}.

\subsection{Continuum limit and spatial slicing}

The resonant $^{57}$Fe layer is treated in the continuum limit by representing discrete nuclei by a continuous density $\rho(\mathbf r)$ and corresponding local transition operators. This description is appropriate whenever the observables of interest vary on length scales much larger than the mean inter-nuclear spacing, such that only coarse-grained spatial averages enter the waveguide-QED dynamics. Numerically, this continuum description is implemented by partitioning the layer into a finite number of slices along the propagation direction, each assumed to be sufficiently thin that the local field and coherence vary weakly across the slice. The guided X-ray mode mediates long-range coherent (\(J\)) and dissipative (\(\Gamma\)) couplings between slices via the electromagnetic Green's function. Each slice is driven by the local field, comprising the incident and retarded fields scattered from all other slices. Related mesoscopic treatments of spatially localized emitter ensembles in waveguide QED have been developed in Ref.~\cite{ruks2023}, where many-emitter systems are likewise reduced to effective continuum descriptions.

For numerical implementation, the interval occupied by the resonant layer is partitioned into $N_x (=100)$ slices of width $\Delta x (=0.5~\mu m)$, centered at positions $x_j$.
Within each slice, the density and nuclear observables are taken to be approximately constant,
\begin{equation}
\rho_N(x)\,\hat{\sigma}_{ij}(x,t)
\;\approx\;
\rho_j\,\hat{\sigma}^{(j)}_{ij}(t),
\quad x\in[x_j-\Delta x/2,x_j+\Delta x/2].
\end{equation}
The continuum integrals appearing in the field equation and in the collective coupling terms are then approximated by sums. In the limit $\Delta x\rightarrow 0$ (at fixed length), this approximation converges to the continuum model. We verified convergence of the computed observables with respect to the number of spatial slices: further refinement does not change $n_e(t)$, $n_{e,\max}$ within numerical tolerance.

Approximations summarized above are controlled by two parameters:
(i) the ratio of the slice size to the characteristic spatial variation scale of the guided field and coherence and (ii) the retardation ratio $\tau_{\mathrm{ret}}/T_1$.
Within the parameter regime considered, they are small, justifying the continuum limit and negligible retardation time approximations.

\subsection{Nonlinear excitations}

Previous treatments of nuclear scattering inside a waveguide using hard X-ray laser were in the regime where the nuclear ensemble responds linearly to the incident field, and the electromagnetic Green tensor enters only as a perturbative correction to the spontaneous emission rate or as a single-pass scattering kernel. The optical Bloch equations are solved under the assumption of a prescribed driving field, without accounting for back-action from the induced nuclear polarization~\cite{Lohse2024}. However, QB operation requires charging protocols reaching finite excitation fractions, with $\langle \hat\sigma_{ee}(x,t)\rangle=O(0.1)$ or larger, so that saturation and nonlinear feedback become relevant and thus need to account for back-action from the induced nuclear polarization. The incident pulse not only excites the nuclear ensemble but also builds up the gradual coherence between the nuclei, as described by Eq.~\eqref{eq:nucl}, over wide timescale evolution: starting from the Gaussian pulse of picoseconds width to the relaxation scale of $\sim 10^2$ ns. As the coherence grows, the scattered field increases and feeds back into the total field via Eq.~\eqref{eq:tfldS}, reshaping the drive and modifying subsequent absorption. The combination of strong driving and long relaxation/dephasing times thus opens access to collective nuclear-exciton dynamics at hard X-ray energies beyond linear response.

We therefore propagate the full optical Bloch dynamics across the distributed ensemble, evolving both the population $\hat\sigma_{ee}(x,t)$ and coherence $\hat\sigma_{ge}(x,t)$ self-consistently under the local field, leading to pronounced temporal signatures, including pulse reshaping and coherent population dynamics that cannot be captured by single-pass scattering models~\cite{Andrejic2023,Sheremet2021}. This input--output/Green’s-function formulation is not restricted to linear response and remains applicable in the nonlinear regime provided the underlying Markov and rotating-wave approximations hold on the scale of the narrow nuclear resonance and the relevant waveguide bandwidth~\cite{Lalumiere2013,Sheremet2021}.

The feasibility and requirements for entering nonlinear nuclear excitation regimes in structured photonic environments have recently been quantified for thin-film X-ray cavities driven by focused pulses~\cite{Lentrodt2024}, motivating the present waveform-engineering approach in the waveguide setting. To suppress preparation-stage radiative loss and enhance absorption, we employ engineered coherent driving. In guided one-dimensional geometries, absorption is highly sensitive to the temporal mode of the incident field, so a naive Gaussian drive is generally not optimal for maximizing excitation in an open radiative system~\cite{Chen2011}.

\begin{figure}[!htbp]
    \centering
    \includegraphics[width=0.45\textwidth]{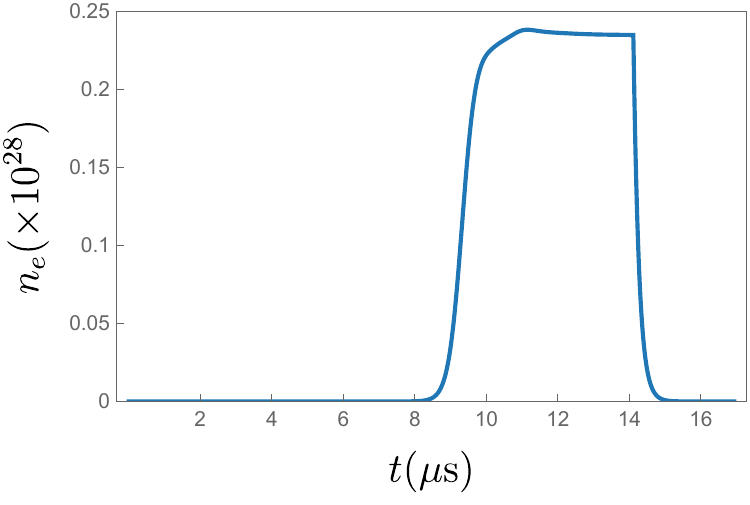}
    \caption{ Total nuclear excitation $n_e(t)$ for a number density $n=10^{28}$ under the self-consistent drive, including the relaxation after the drive is switched off. The excitation rises rapidly during and reaches a quasi-steady plateau while the drive is applied and then decays after the drive ends. This decay rate is close to the natural decay rate, indicating that no substantial collective enhancement of the decay occurs once the driving field is removed.}
    \label{fig:s1}
\end{figure}


\subsection{Fixed incident energy per nucleus}

For each ensemble density $n$, we optimize the incident temporal mode under a constraint of fixed resource per nucleus, $\int_0^{t_{\mathrm{End}}}|\Omega_{\mathrm{in}}(t)|^2dt/n=\mathrm{const.}$, and obtain the corresponding self-consistent waveform over the interval $[0,t_{\mathrm{End}}]$. This optimized waveform is then used to excite the nuclear ensemble. Figure~\ref{fig:s1} shows the resulting nuclear excitation $n_e(t)$, followed by the subsequent relaxation after the drive is switched off at $t_{\mathrm{End}}$.

In the linear regime, the excitation is approximately proportional to the nuclei to photon ratio. However, the proportionality starts to breaks on entering the nonlinear regime and there will ultimately be saturation such that the excitation is independent of the nuclei to photon ratio.

\section{Long-retention nuclear QBs.}

Table~\ref{t:isomers} lists representative nuclear isomers that can be modeled
as effective three-level systems with a metastable storage state, making them
potential candidates for long-retention nuclear QBs.

\begin{table}[th]
\centering
\caption{Selected nuclear-isomers with isomer excitation energy $E_{\rm I}$, half-life $T_{1/2}$ and spin level for long-retention nuclear QBs.}
\label{t:isomers}
\begin{tabular}{|l|c|c|c|}
\hline
Isomer & $E_{\rm I}$ & $T_{1/2}$ & Spin state \\
\hline
$^{178\mathrm{m}2}$Hf & $2.4$ MeV & $31$ y & $I_e=16; I_g=0$ \\
$^{180\mathrm{m}}$Ta & 75 keV & $>10^{15}$ y & $I_e=9; I_g=1$ \\
$^{242\mathrm{m}}$Am & 48.6 keV & $143$ y & $I_e=5; I_g=1$ \\
$^{97\mathrm{m}}$Tc & $143$ keV & $6$ h & $I_e=1/2; I_g=9/2$ \\
$^{93\mathrm{m}}$Mo & $2.4$ MeV & $6.8$ h & $I_e=21/2; I_g=5/2$ \\
$^{129\mathrm{m}}$Sb & $1.8$ MeV & $17.7$ m & $I_e=19/2; I_g=7/2$ \\
\hline
\end{tabular}

\end{table}

\putbib[origin]
\end{bibunit}

\end{document}